\documentclass[rnaas]{aastex631}

\usepackage{hyperref}

\begin{document}

\title{Discovery of Raman-Scattered \ion{C}{2} Lines in V366 Carinae with GHOST/Gemini South}

\author[0000-0003-2530-3000]{Jeong-Eun Heo}
\affiliation{Gemini Observatory/NSF NOIRLab, Casilla 603, La Serena, Chile}

\author[0009-0009-7838-7771]{Miji Jeong}
\affiliation{Department of Astronomy, Space Science, and Geology, Chungnam National University, Daejeon 34134, Republic of Korea}



\begin{abstract}
We report the detection of Raman-scattered \ion{C}{2} lines at 7023 and 7054~\AA\ in the symbiotic star V366~Carinae using GHOST at  Gemini South. These faint features, originating from the \ion{C}{2} doublet at $\lambda\lambda$ 1036 and 1037~\AA, are rare and have been detected in only two other symbiotic stars: V1016~Cygni and RR~Telescopii. Our findings showcase the exceptional sensitivity of GHOST to detect subtle spectral features and open the door to comparative studies of Raman-scattered \ion{C}{2} features across these systems. 
\end{abstract}




\section{Introduction} \label{sec:intro}
Symbiotic stars are fascinating binary systems composed of a hot compact star and a cool giant star, interacting through mass transfer and exhibiting emissions across the electromagnetic spectrum. Among the various phenomena observed in these systems, Raman scattering - an inelastic scattering of ultraviolet (UV) photons by neutral hydrogen (\ion{H}{1}) - produces unique optical emission features \citep{schmid1989}. 

The Raman \ion{O}{6} features at 6825 and 7082~\AA\ are among the most well-studied examples, often displaying multiple-peaked profiles that reflect the complex kinematics and scattering geometries in the \ion{H}{1} region. These features have provided critical insights into the interactions between the UV radiation field and the surrounding \ion{H}{1} region, establishing Raman scattering as a powerful tool for studying symbiotic systems \citep{lee2022,heo2021}.

Raman-scattered \ion{C}{2} features at 7023 and 7054~\AA\ arise from the \ion{C}{2} doublet at $\lambda\lambda$ 1036 and 1037~\AA. Although these lines were theoretically predicted more than three decades ago \citep{nuss1989}, they have only been observed in two symbiotic stars to date: V1016~Cygni \citep{schild1996} and RR~Telescopii \citep{heo2019}. Their rarity makes their detection particularly valuable, as Raman \ion{C}{2} lines serve as a complementary diagnostic to Raman \ion{O}{6}, offering new perspectives on Raman scattering mechanisms as well as the dynamics and physical conditions of these intriguing systems.

\section{Observation and Data Reduction}
We observed the symbiotic star V366~Car with the Gemini High-resolution Optical SpecTrograph \cite[GHOST;][]{mc2024,kalari2024} on the 8.1m Gemini South telescope at Cerro Pachón, Chile. The observations were part of the System Verification program (GS-2023A-SV-103\footnote{\url{https://archive.gemini.edu/searchform/GS-2023A-SV-103/}}) conducted on 2023 May 16 UT.

We used the standard-resolution mode with 1$\times$2 (spectral $\times$ spatial) binning, which delivers a spectral resolving power R $\sim$ 56,000. The data set includes three exposures, each with an integration time of 150 seconds. Observing conditions were clear, with the average seeing of $\sim$1.0\arcsec. For flux calibration, we observed the spectrophotometric standard LTT~4816 on the same night with the same instrument setup. 

Data reduction was carried out using \texttt{GHOSTDR} v1.0.0 \citep{ireland2018,hayes2022}, a customized version of the Gemini Observatory’s \texttt{DRAGONS}\footnote{\url{https://github.com/GeminiDRSoftware/DRAGONS}} data-reduction platform \citep{labrie2023}, developed during the commissioning of GHOST. The reduction process included bias and flat-field corrections, wavelength calibration, sky subtraction, barycentric correction, order combination, and flux calibration. 

{
\centering
\begin{figure}[htp]
\epsscale{0.75}
\plotone{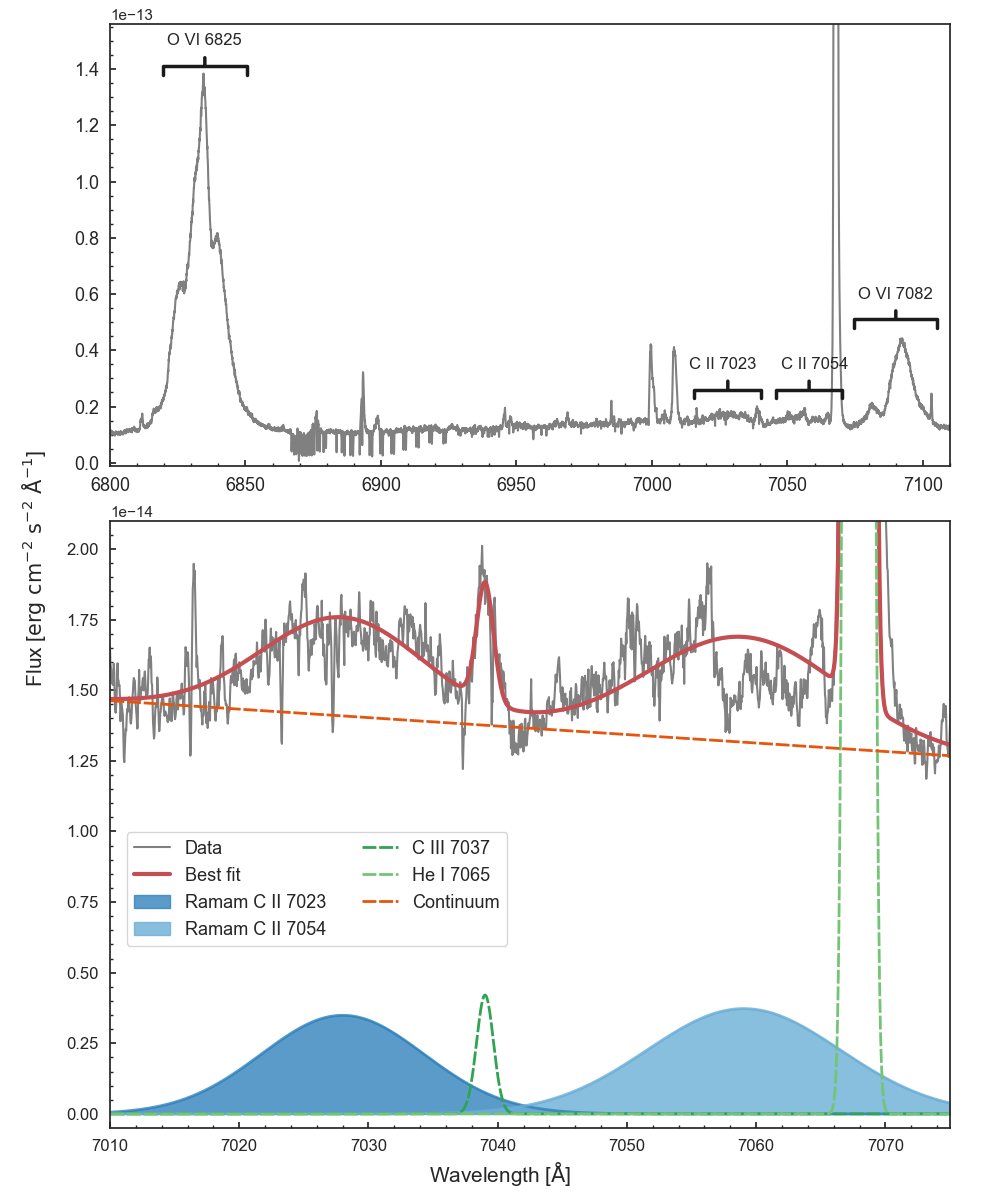}
\caption{\textit{Top:} The GHOST spectrum of V366~Car, showing the Raman \ion{O}{6} and \ion{C}{2} features. \textit{Bottom:} A zoom-in view of the Raman-scattered \ion{C}{2} features, along with the fitting results. The blue and light-blue shaded regions represent the best-fit single Gaussian profiles for the Raman \ion{C}{2} features. \label{fig:v366car}}
\end{figure}
}

\section{Results}
Fig.~\ref{fig:v366car} presents a portion of the GHOST spectrum of V366~Car. The GHOST data reveals strong Raman \ion{O}{6} features at 6825 and 7082~\AA. The highlight of this observation is the serendipitous discovery of Raman \ion{C}{2} $\lambda\lambda$ 1036 and 1037 features at 7023 and 7054~\AA.
A key difference between Raman \ion{O}{6} and Raman \ion{C}{2} features is their profile shapes. While Raman \ion{O}{6} features exhibit complex multiple-peaked profiles, the Raman \ion{C}{2} features display simpler, single-peaked profiles. This contrast likely reflects the differences in the scattering regions or the kinematics of the UV-emitting regions for these two sets of features. Raman \ion{O}{6} features are thought to trace more complex dynamics, such as binary motion or wind interactions, whereas the Raman \ion{C}{2} features may originate from more stable or localized scattering regions.

To identify the Raman \ion{C}{2} features, we first fit the continuum with a linear function and deblended the nearby \ion{C}{3} $\lambda$ 7037 and \ion{He}{1} $\lambda$ 7065 emission lines using single Gaussian functions. The Raman \ion{C}{2} features were then modeled using the single Gaussian profiles, with the blue and light-blue shaded regions in Fig.~\ref{fig:v366car} representing the best fits. 

The line center wavelengths for the Raman \ion{C}{2} features were measured at 7027.99~\AA\ and 7059.03~\AA, with FWHM of 14.81~\AA\ and 17.62~\AA, respectively. 
The FWHM values are slightly broader than those reported in V1016~Cyg and RR~Tel, where FWHM values of 12.5~\AA\ and 11.3~\AA, respectively. These broader profiles in V366~Car may indicate enhanced turbulence or velocity dispersion for its scattering region.

The total line fluxes are measured as $F(7023) = 5.50 \times 10^{-14} \, \text{erg} \, \text{cm}^{-2} \, \text{s}^{-1}$ and $F(7054) = 6.99 \times 10^{-14} \, \text{erg} \, \text{cm}^{-2} \, \text{s}^{-1}$.
The nearly identical flux values of the two Raman \ion{C}{2} features are consistent with theoretical expectations for the \ion{C}{2} $\lambda\lambda$ 1036,1037~\AA\ doublet, which has similar intrinsic strengths under optically thick conditions.

\section{Summary and Discussion}
The detection of Raman \ion{C}{2} lines in V366~Car adds a third example to the exclusive group of symbiotic stars exhibiting these rare features alongside V1016~Cyg and RR~Tel. The discovery of these rare Raman \ion{C}{2} features opens new opportunities for exploring symbiotic systems, particularly when combined with Raman \ion{O}{6} analysis.

Comprehensive studies of Raman-scattering in symbiotic systems require detailed radiative transfer simulations. Modeling the UV radiation field, H I distribution, and the physical conditions of the scattering regions may help constrain the spatial extent and kinematic properties of the regions responsible for Raman scattering. Furthermore, such studies will provide a framework to explore different types of Raman-scattered features and their sensitivity to the geometry and dynamics of the system.

Future observations of similar systems will also be crucial for establishing whether the presence of Raman \ion{C}{2} features correlates with specific evolutionary stages or physical conditions in symbiotic stars.


\section{Acknowledgments}
\begin{acknowledgments}
Based on observations obtained under Program ID GS-2023A-SV-103 at the international Gemini Observatory, a program of NSF NOIRLab, which is managed by AURA under a cooperative agreement with the U.S. National Science Foundation on behalf of the Gemini Observatory partnership.

GHOST was built by a collaboration between Australian Astronomical Optics at Macquarie University, National Research Council Herzberg of Canada, and the Australian National University, and funded by the International Gemini partnership. The instrument scientist is Dr. Alan McConnachie at NRC, and the instrument team is also led by Dr. Gordon Robertson (at AAO), and Dr. Michael Ireland (at ANU).

The authors would like to acknowledge the contributions of the GHOST instrument build team, the Gemini GHOST instrument team, the full SV team, and the rest of the Gemini operations team that were involved in making the SV observations a success.
\end{acknowledgments}

%

\vspace{5mm}
\facilities{Gemini:South (GHOST)}





\begin{thebibliography}{}
\bibitem[Hayes et al.(2022)]{hayes2022} Hayes, C.~R., Waller, F., Ireland, M., et al.\ 2022, \procspie, 12184, 121846H
\bibitem[Heo et al.(2019)]{heo2019} Heo, J.-E., Lee, H.-W., Angeloni, R., et al.\ 2019, Why Galaxies Care About AGB Stars: A Continuing Challenge through Cosmic Time, 343, 416
\bibitem[Heo et al.(2021)]{heo2021} Heo, J.-E., Lee, H.-W., Angeloni, R., et al.\ 2021, \apj, 915, 105
\bibitem[Ireland et al.(2018)]{ireland2018} Ireland, M.~J., White, M., Bento, J.~P., et al.\ 2018, \procspie, 10707, 1070735
\bibitem[Kalari et al.(2024)]{kalari2024} Kalari, V.~M., Diaz, R.~J., Robertson, G., et al.\ 2024, \aj, 168, 208
\bibitem[Labrie et al.(2023)]{labrie2023} Labrie, K., Simpson, C., Cardenes, R., et al.\ 2023, Research Notes of the American Astronomical Society, 7, 214
\bibitem[Lee et al.(2022)]{lee2022} Lee, Y.-M., Kim, H., \& Lee, H.-W.\ 2022, \apj, 931, 142. doi:10.3847/1538-4357/ac67d6
\bibitem[McConnachie et al.(2024)]{mc2024} McConnachie, A.~W., Hayes, C.~R., Robertson, J.~G., et al.\ 2024, \pasp, 136, 035001
\bibitem[Nussbaumer et al.(1989)]{nuss1989} Nussbaumer, H., Schmid, H.~M., \& Vogel, M.\ 1989, \aap, 211, L27
\bibitem[Schmid(1989)]{schmid1989} Schmid, H.~M.\ 1989, \aap, 211, L31
\bibitem[Schild \& Schmid(1996)]{schild1996} Schild, H. \& Schmid, H.~M.\ 1996, \aap, 310, 211
\end{thebibliography}



\end{document}